\documentclass{bvm} 

\begin{document}

\newcommand{\bvmyear}{2026}

\selectlanguage{english} 

\title{Exploring General-Purpose Autonomous Multimodal Agents for Pathology Report Generation}


\titlerunning{BVM \bvmyear}

\author{
	\fname{Marc} \lname{Aubreville} \inst{1} \isResponsibleAuthor,    
	\fname{Taryn A.} \lname{Donovan} \inst{2} \affiliation{Schwarzman Animal Medical Center, New York},
    \fname{Christof A.} \lname{Bertram} \inst{3} \affiliation{University of Veterinary Medicine, Vienna}
}

\authorrunning{Aubreville et al.}

\institute{
\inst{1} Flensburg University of Applied Sciences, Flensburg, Germany\\
\inst{2} Schwarzman Animal Medical Center, New York, USA \\
\inst{3} University of Veterinary Medicine Vienna, Vienna, Austria \\
}

\email{marc.aubreville@hs-flensburg.de}

\maketitle

\begin{abstract}
Recent advances in agentic artificial intelligence, i.e. systems capable of autonomous perception, reasoning, and tool use, offer new opportunities for digital pathology. In this pilot study, we evaluate whether two agentic multimodal AI systems (OpenAI’s ChatGPT 5.0 in agentic mode, and H Company's Surfer) can autonomously navigate, describe, and interpret histopathologic features in digitized tissue slides on a slide viewing platform. A set of 35 veterinary pathology cases, curated for training purposes, was used as the test dataset. The agent was tasked with autonomously exploring whole-slide images using a web-based slide viewer, identifying salient tissue structures, generating descriptive summaries, and proposing provisional diagnoses. We fed different prompts to explore three scenarios: 1) analysis without knowledge of the signalment, 2) analysis with organ and species provided, and 3) diagnosis based on a morphological description provided. All outputs were reviewed and validated by a board-certified pathologist for accuracy and diagnostic consistency. We further tasked another board-certified pathologist with the same task to establish a baseline. We found the systems to yield accurate diagnoses in up to 28.6\% of cases with only images, signalment and organ provided, and up to 68.6\% when a morphological description was provided. With only the WSI provided, the models were only correct in up to 5.7\% of cases. The human expert, on the other hand, achieved 85.7\% diagnostic accuracy with only a single WSI, and 88.6\% when also signalment and organ was provided.

The study demonstrates that while the agentic AI system can meaningfully engage with web-based slide viewing software to assess complex visual pathology data and produce contextually aligned feature descriptions, diagnostic precision remains limited compared with a human expert.
\end{abstract}

\section{Introduction}
The establishment of digital pathology workflows in labs allows for the use of artificial intelligence (AI) based solutions to be embedded into the diagnostic process. While current research shows that methods based on deep learning models can achieve performance on the same level of or even exceeding pathology experts, the tasks considered are typically narrowly defined tasks, such as mitotic figure identification \cite{aubreville2024domain}, cell or tissue segmentation \cite{ma2024multimodality}, or tumor grading \cite{bulten2022artificial}. 

Report generation is the final step in the pathological diagnosis and involves a multitude of pattern recognition steps, in combination with reasoning and navigation on the pathology slide(s) at various magnifications. Given the complexity of this task, it could be assumed that this is an unachievable task for current artificial intelligence systems. Yet, recent models released in the field of agentic artificial intelligence have demonstrated emerging capabilities for autonomous reasoning, multimodal perception, and iterative goal-directed behavior. In particular, the use of large vision and language models, in combination with tool use, has enabled browser-based navigation of websites, supported by planning and reasoning of workflows. 
Large-language models models have gained remarkable capabilities across a wide range of domains, and while domain-specific training does improve on the results, even general-purpose models can be applied successfully -- albeit with limitations -- to specialized fields such as medicine~\cite{saab2024capabilities}.

This raises an intriguing question: Can such general-purpose AI agents emulate aspects of the diagnostic workflow, specifically, the exploration, recognition, and verbal synthesis steps performed by pathologists? 

In this study, we investigate the feasibility of deploying an agentic AI system within an open source digital slide viewing environment to autonomously perform slide navigation, tissue feature description, and preliminary diagnostic reasoning. To the best of our knowledge, this is the first attempt at testing the capability of commercially available general-purpose AI tools at this highly specialized task. While we do not advocate for the automation of pathology report generation in real-world clinical settings (given the significant ethical and safety concerns involved), we propose that this task offers a valuable benchmark for assessing the emerging capabilities of agentic vision–language models.

\begin{figure}
\includegraphics[height=3.6cm]{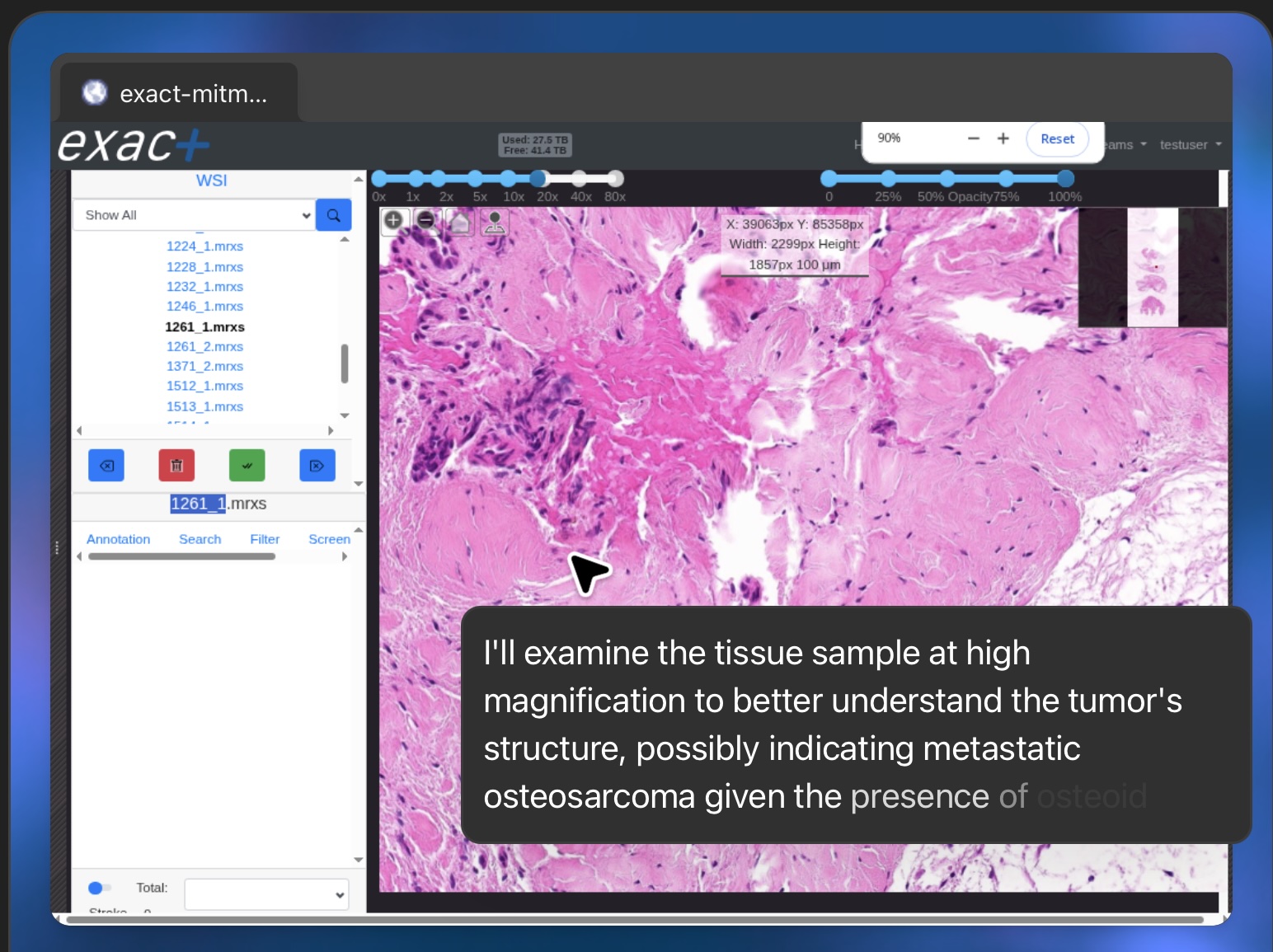}
\includegraphics[height=3.6cm]{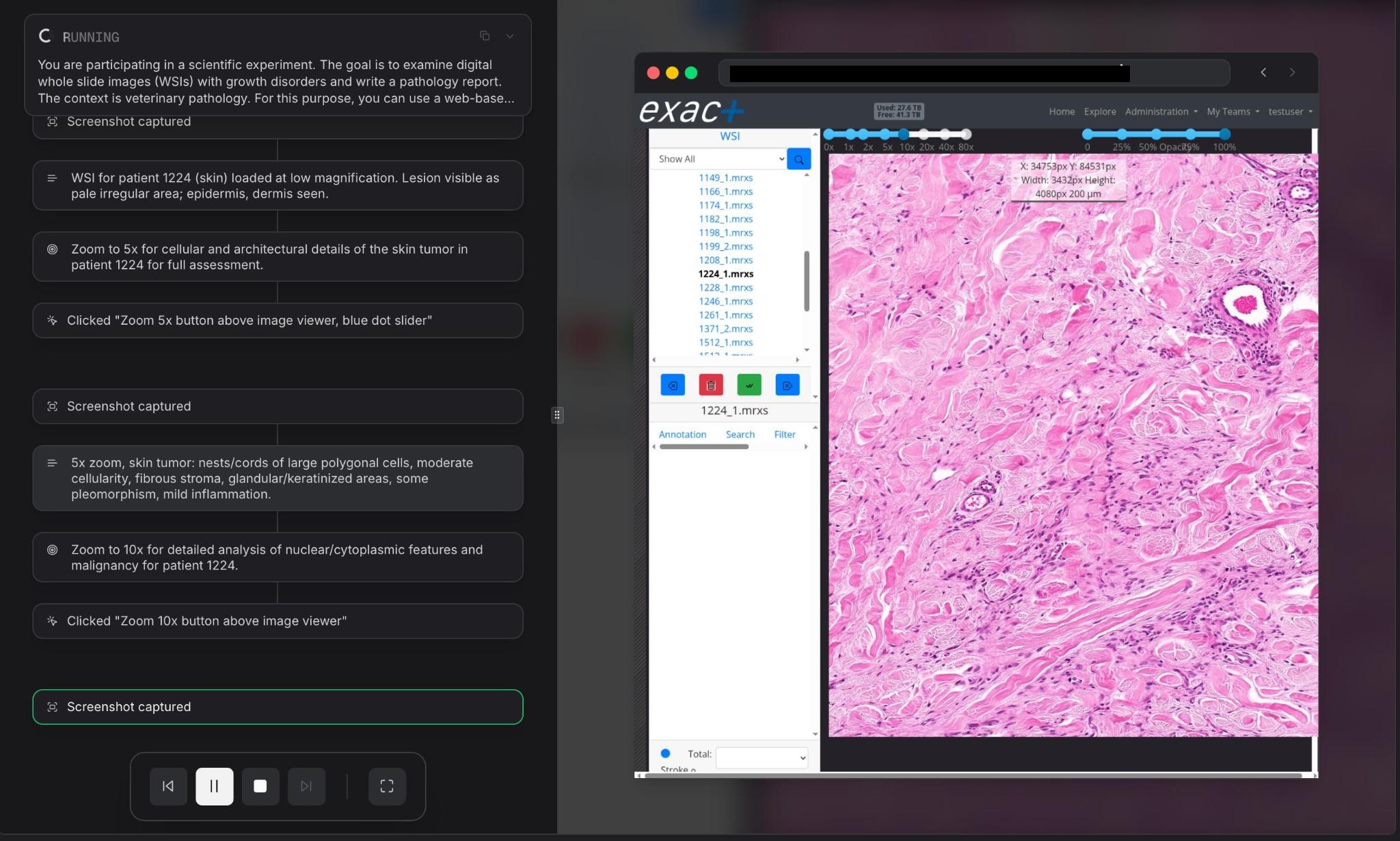}
\caption{Screenshots of the experiment, showing ChatGPT 5.0 (OpenAI, left panel) and Surfer (H Company, right panel). Both investigated agentic AI frameworks were able to successfully navigate the slides and complete the tasks.}
\end{figure}

\section{Methods}
We prepared an online slide viewing platform~\cite{marzahl2021exact} with a dataset of 35 cases from veterinary pathology, retrieved from the diagnostic archive of the University of *anonymized*. These cases were part of a panel of educational cases of neoplastic and tumor-like lesions for the training of students and residents in veterinary pathology. Naturally, this means that the cases were selected to be interesting, rather than representative of the general sample distribution, and had an overall good quality. Furthermore, it means that these cases were selected to be diagnostically characteristic, which facilitates the evaluation in this experiment. For each case, we have the following data records:
\begin{enumerate}
    \item A single whole slide image (WSI) for each case using hematoxylin and eosin stain. 
    \item The organ the specimen was retrieved from.
    \item The signalment, i.e., the animal's species and breed, as well as age and sex.
    \item The complete morphologic description by a board-certified pathologist, following a structured catalog, but given in plain text. The morphological description contains all features required to infer the diagnosis.
    \item The diagnosis, including information which type of growth disorder (benign or malignant tumor, hyperplasia, ...) was determined by a board-certified veterinary pathologist.
\end{enumerate}
Out of the 35 cases, 28 were from dogs, six from cats, one from a guinea pig, and one from a horse. Regarding organs, the largest organ represented was the skin~(13/35), reflecting the high sampling frequency of this organ in domestic animal diagnostic pathology. Furthermore, there were samples from adrenal gland (N=4), spleen (N=4),  mammary gland / breast (N=2),  eye (N=2), intestine (N=2), joint (N=2), oral mucosa (N=2), uterus (N=2), vulva (N=1), and central nervous system (N=1), covering a wide range of organs. 
From this, we define three scenarios with progressively more conditioning or context provided to the model:
\begin{itemize}
    \item Scenario A – WSI only: The model received only the WSI and was tasked with inferring both the organ of origin and the diagnosis solely from the image content.
    \item Scenario B – WSI + brief case description: The model was provided with the WSI along with a concise case description, including signalment (species, breed, sex, age, and neuter status) and the organ of origin.
    \item Scenario C – Morphologic description only: The model received only a textual morphologic description. Because the complete morphology was already represented in the description, the WSI was not provided to accelerate inference.
    
\end{itemize}

From the dataset, we derived prompts, requiring the model to provide a structured JSON object per case. In this, we requested the model to provide the organ, the diagnosis, as well as multiple morphologic descriptions to provide the model with an incentive to investigate the specimen thoroughly. This list consists of shape (of mass), demarcation (of tumor/surrounding tissue), invasion, cellularity (of mass), growth pattern, stroma and matrix (amount and type), other subgross features (such as hemorrhage or superficial ulceration), tumor cells (size, shape, cell borders, differentiation), tumor cytoplasm (amount, character, content), tumor nuclei (shape, location, chromatin distribution), tumor nucleoli (number and size), malignancy criteria (nuclear pleomorphism and mitotic activity), and other tumor features (necrosis, inflammation, or multinucleation). The detailed prompts and example videos are provided in our github repository\footnote{\url{https://github.com/DeepMicroscopy/AgenticPathologyReports/}}. We used batches of three to expedite the experiment, striking a compromise between potentially running into the context limits of the model (as a result of having too many cases) and limiting the time required for the initial setup of the model and the log-in process. 
In the slide viewing platform, a single dataset was created to be visible for the client. We specifically instructed the model with hints on navigation on the slide and in the software, which we found to be beneficial in early experiments. 

We surveyed two agentic multimodal AI solutions that allow for the direct formulation of tasks to be executed on websites: OpenAI's ChatGPT 5.0 in agentic mode, as well as H Company's Surfer \cite{andreux2025surfer}, which is built on the publicly available Holo1 v1.5 model \cite{holo15_68c1a5736e8583a309d23d9b}. Both frameworks were utilized using their respective web user interfaces. If available, caches and memories were turned off for the experiment. After retrieving results from each suite, a board-certified veterinary pathologist evaluated all responses for accuracy. We evaluated the detailed textual morphologic descriptions only for the leading model, as the second-best model yielded notably subpar performance, making an in-depth analysis of its morphologic descriptions less informative. As a benchmark for the AI agents, we asked a second board-certified pathologist to evaluate the WSIs at two time points with the same (limited) information available as the AI agents. 

\section{Results}

Our results, shown in Table \ref{results_table}, indicate that both frameworks had considerable difficulties performing this highly complex task, as opposed to a pathologists who made the correct diagnosis in >85\% of the cases. While we did not observe major obstacles involving the navigation within the slide viewing software, the correctness of the models was generally low. However, we observed that the models did, in many cases, not visually inspect the most relevant parts of the slide, which contributes to a low performance, since slide areas containing potentially important visual features were never inspected. This is reflected by numerous erroneous descriptions of ChatGPT in 336 of 455 features (73.85\%, see Table \ref{tab:detailed_results} for details). An important observation was that the models never avoided providing a diagnosis or detailed morphologic descriptions, but delivered an answer that might be probable based upon the information provided (i.e. signalment and organ), listing common tumors for a given species and organ, but not the actual tumor type depicted in the case being examined. 
This is also reflected by the increased rate of correct diagnoses and descriptions (error rate of ChatGPT: 270 of 455 features, 59.3\%) once the signalment and organ were provided to the model. 

In the identification of the corresponding organ, we saw a considerable difference between both frameworks, with the ChatGPT framework having a higher performance across all conditions. We also found that this model was better at contextualizing the provided brief case description in scenario B.
If a morphological description was provided to the models, the likelihood of a correct diagnosis of both models was much higher (see Table \ref{results_table}). If the organ was provided; either as part of the brief case description (scenario B) or as part of a verbal morphological description, the models had almost no difficulty in identifying the corresponding organ.  

\begin{table}
\caption{Results of our experiment, evaluated by a board-certified pathologist. Both evaluated frameworks benefit from knowing the organ and signalment, and more so by a morphological description.}
\resizebox{\linewidth}{!}{\begin{tabular}{p{3.5cm}lrrrr}
\hline
Framework & Trial & Organ correct & \multicolumn{3}{c}{Diagnosis}  \\
 & & & correct & wrong dignity & wrong subtype \\
\hline
\multirow{3}{*}{ChatGPT 5.0 (OpenAI)} & A: Only WSI available & 28.6\% & 5.7\% & 5.7\% & 2.9\% \\
 & B: WSI + signalment + organ given & 100.0\% & 28.6\% & 2.9\% & 2.9\% \\
 & C: Morphological description given & 100.0\% & 68.6\% & 8.6\% & 2.9\% \\
 \hline
\multirow{3}{*}{Surfer (H Company)} & A: Only WSI available & 11.4\% & 2.9\% & 0.0\% & 5.7\% \\
 & B: WSI + signalment + organ given & 97.1\% & 8.6\% & 2.9\% & 17.1\% \\
  & C: Morphological description given & 97.1\% & 68.6\% & 5.7\% & 17.1\% \\
\hline
\multirow{2}{*}{Pathologist} & A: Only WSI available & 91.5\% & 85.7\% & 5.7\% & 0\% \\
 & B: WSI + signalment + organ given & N/A & 88.6\% & 5.7\% & 0\% 
   \\
\hline
\end{tabular}}
\label{results_table}
\end{table}

\begin{table}[ht]
\caption{Accuracy of individual morphologic descriptions given by the leading model (ChatGPT 5.0), as evaluated by a board-certified pathologist.}
\resizebox{\linewidth}{!}{\begin{tabular}{llllllll}
\hline
information provided & shape of mass & demarcation & invasion & cellularity & growth pattern & stroma and matrix & other subgross features  \\
\hline
only WSI & 20.0\% & 11.4\% & 20.0\% & 51.4\% & 14.3\% & 22.9\% & 14.3\%  \\
WSI + signalment + organ & 40.0\% & 40.0\% & 51.4\% & 68.6\% & 45.7\% & 31.4\% & 22.9\%   \\
\hline
information provided &  tumor cells & tumor cytoplasm & tumor nuclei & tumor nucleoli & malignancy criteria & other tumor features \\
\hline
only WSI & 42.9\% & 14.3\% & 48.6\% & 34.3\% & 34.3\% & 11.4\% \\
WSI + signalment + organ & 45.7\% & 31.4\% & 51.4\% & 34.3\% & 45.7\% & 20.0\%
\end{tabular}}
\label{tab:detailed_results}
\end{table}

\section{Discussion}
The observations in our study align with prior observations of model hallucination in large (vision and language) models being more likely in the long tail of the data distribution \cite{huang2025survey}, which this use case is certainly representative of. Given the improved rate of diagnostic success in the third scenario (with morphological description given), we can hypothesize that the most challenging part of this task might have been the targeted navigation, in search for diagnostically relevant features, and the extraction and description thereof. This would also correlate with expectations, as the training data for vision-language models (VLMs) is likely to be underrepresented with examples executing diagnostic workflows in pathology, while textual descriptions linking morphological descriptions and diagnoses are likely to be part of the scientific literature that was used in training the models. Although pathology-specific VLMs have recently been introduced~\cite{zhang2025pathoagenticragmultimodalagenticretrievalaugmented}, this study focused on evaluating general-purpose VLMs due to their broad accessibility and frequent use. In our observation, both students and medical practitioners increasingly rely on these general models as informal knowledge resources for medical inquiries. While the evaluated VLMs demonstrated the ability to employ appropriate medical terminology --- potentially conveying an impression of domain competence --- our findings reveal a considerable rate of error in diagnostic interpretation and in the description of histopathological images. These results underscore the importance of exercising caution when using VLMs in educational or clinical contexts.

We acknowledge that our study was limited to only two frameworks that provided integration of web-based browsing with agentic models, and that other, and in particular specialized, models and/or pipelines \cite{zhang2025pathoagenticragmultimodalagenticretrievalaugmented} might be more suitable for this task. However, our primary objective was not to benchmark performance across the full spectrum of available models, but rather to explore the feasibility and qualitative behavior of current general-purpose VLM-based agents when applied to a highly domain-specific diagnostic task.



\printbibliography

\end{document}